# A Theory of Complex Adaptive Learning and a Non-Localized Wave Equation in Quantum Mechanics

**Leilei Shi (石磊磊)**[1,2,*,+], **Xinshuai Guo (郭新帅)**[1,*], **Jiuchang Wei (魏玖长)**[1,*],

**Wei Zhang (张伟)**[2], **Guocheng Wang (王国成)**[3], **Bing-Hong Wang (汪秉宏)**[4,*]

[1]School of Management, University of Science and Technology of China, Hefei 230026, P. R. China
[2]Beijing YourenXiantan Science & Technology Co. Ltd., Beijing 100080, P. R. China
[3]Institute of Quantitative & Technological Economics, Chinese Academy of Social Sciences, Beijing 100732, P. R. China
[4]Department of Modern Physics, University of Science and Technology of China, Hefei 230026, P. R. China
*Corresponding authors (emails: Shileilei8@163.com (Leilei Shi); guoxs@ustc.edu.cn (Xinshuai Guo); weijc@ustc.edu.cn (Jiuchang Wei), bhwang@ustc.edu.cn (Bing-Hong Wang))
+Shi contributed to half of this work, and the others contributed equally.

June 16, 2024

## ABSTRACT

Complex adaptive learning is intelligent. It is adaptive, learns in feedback loops, and generates hidden patterns as many individuals, elements or particles interact in complex adaptive systems (CAS). CAS highlights adaptation in life and lifeless complex systems cutting across all traditional natural and social sciences disciplines. However, discovering a universal law in CAS and understanding the underlying mechanism of distribution formation, such as a non-Gauss distribution in complex quantum entanglement, remains highly challenging. Quantifying the uncertainty of CAS by probability wave functions, the authors explore the inherent logical relationship between Schrödinger's wave equation in quantum mechanics and Shi's trading volume-price wave equation in finance. Subsequently, the authors propose a non-localized wave equation in quantum mechanics if cumulative observable in a time interval represents momentum or momentum force in Skinner-Shi (reinforcement-frequency-interaction) coordinates. It reveals that the invariance of interaction as a universal law exists in quantum mechanics and finance. The theory shows that quantum entanglement is an interactively coherent state instead of a consequence of the superposition of coherent states. As a resource, quantum entanglement is non-separable, steerable, and energy-consumed. The entanglement state has opposite states subject to interaction conservation between the momentum and reversal forces.

**Keywords:** complex adaptive systems, complex adaptive learning, complex quantum systems, non-localized wave equation, interaction conservation, interactively coherent entanglement

**PACS:** 89.75.-k (Complex Systems); 89.65.Gh (Economics, Econophysics, Financial Markets, Business and Management); 03.65.Ud (Entanglement and Quantum Nonlocality)

## Introduction

Complexity sciences are interdisciplinary fields that specialize in studying complexity and complex systems, including lifeless and life systems such as dissipative structures in thermodynamics [1] and cellular structures in biology [2]. The sciences span various research fields investigating the systems behaviors of particles, agents, and organisms from natural to social sciences. Complex systems consist of many interacting elements, particles or individuals, show uncertainty by chance, and emerge multilevel patterns. They have been studied by physicists for centuries [3-7]. They are nonlinear, non-monotonic, and more than the sum of the parts [8], resembling a probability wave in financial markets [9]. They occur almost everywhere, cutting across all traditional natural and social sciences disciplines, including mathematics, physics, chemistry, biology, engineering, psychology, medicine, neuroscience, artificial intelligence, computer science, economics, finance, management, Etc. In 1992, Holland proposed a complex adaptive systems (CAS) framework to study, compute, and simulate complex systems [10]. It has a standard kernel extracted from complex systems, which involves many components, units or agents that adapt as they interact. The framework combines the interaction of elements, particles, or individuals with mutual causation in feedback loops, reflecting common adaptive characteristics in many complex systems. It has a profound impact on complexity sciences. Complex adaptive systems (CAS) are groups of adaptive elements, particles, or agents that interact in interdependent ways to produce system-wide patterns. They highlight adaptation in the framework for studying, explaining, and understanding complex systems in which the semi-autonomous elements, particles, or agents collectively combine to form emergent and global-level interaction properties [11]. Extending the scope of CAS by behavior analysis in psychology, we introduce complex adaptive learning [12-14]. Complex adaptive learning is intelligent. It is adaptive, learns in feedback loops, and generates hidden patterns over a reinforcement range as elements, particles, or agents interact. The elements, particles, or agents can be nearly anything, from brain cells or viruses to ants or bees, to water particles in a weather pattern or iron atoms in a spin glass, Etc. [4] [5]. The reinforcement can be whatever thing to which interacting elements, particles, or agents are sensitive, such as prices to traders in finance, foods to ant colonies in behavior analysis, proteins to neurons in biology, magnetic fields to iron atoms in condensed matter physics, lasers to photons in complex quantum systems, Etc. Complexity in systems has attracted significant attention from scientists. Stephen Hawking predicted, "I think the next century (the 21$^{st}$ century) will be the century of complexity" [15].

In the 21st century, a growing number of pioneering scientists believe that complex systems spanning different disciplines have commonalities and are driven by the exact underlying mechanism [6] [11] [16-18]. Exploring a universal law and proposing a theory in CAS has become a desirable goal [19]. It is a rare opportunity once in a century in the history of sciences. However, discovering a universal law and understanding the underlying mechanism of pattern formation in CAS has encountered many challenges, such as the intangibility in interactions, the uncertain, nonlinear, and non-monotonic variables, the interdependence in feedback loops, and the estimation of unknown parameters in the models [20-22].

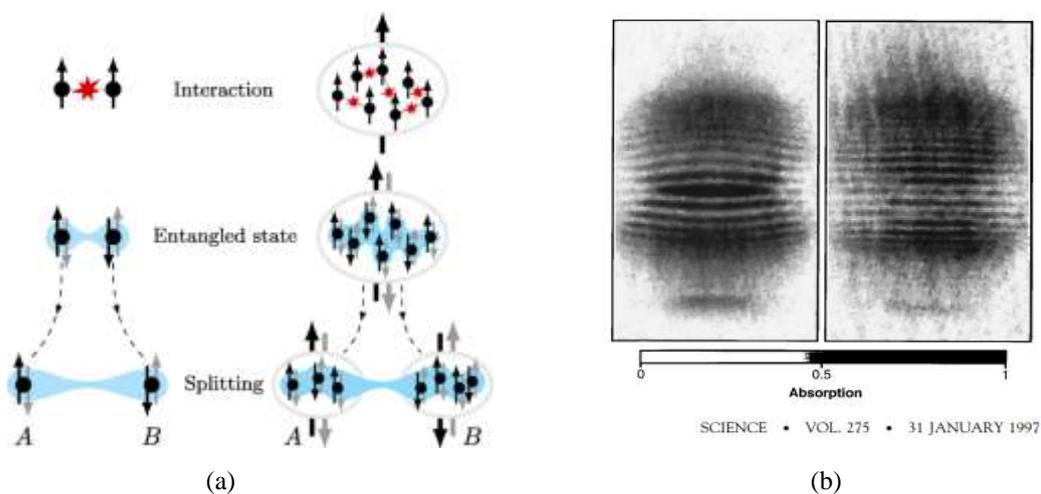

(a) (b)

Fig. 1: Quantum entanglement

Note: (a) Schematic of an EPR experiment with two particles (left) and with two many-particle systems (right) [26]; (b) Interference pattern of two expanding condensates observed after 40 ms time-of-flight for two different powers of the argon ion laser-light sheet (raw-data images) [27].

This paper attempts to extract a universal law of complex adaptive learning in CAS from a trading volume-price probability wave equation in the financial markets since the markets are typical complex adaptive systems, generate massive data every trading day, and emerge many observable phenomena for study [12]. We assume that the law applies to various disciplinary fields in complex adaptive systems and contributes to understanding the underlying mechanism of quantum entanglement in Einstein-Podolsky-Rosen (EPR) experiments (shown in Fig. 1) [23] [24], such as non-Gauss distributions in complex quantum entanglement [25], multiparticle entanglement distributions in Bose-Einstein condensates (BEC) [26], and observation of interference between two freely expanding Bose-Einstein condensates (BEC) [27], Etc. Liu et al. [28] assume that a nonlinear item represents entanglement in the one-dimensional nonlinear Schrödinger or Gross-Pitaevskii equation. They study nonlinear effects in the interference of BEC using exact solutions of the one-dimensional nonlinear Schrödinger equation. With the inverse scattering method, they find the interference pattern as a scattering problem with the linear Schrödinger equation, whose potential is profiled by the initial density distribution of the condensates. Recently, Zhang et al. [29] report the formation of density patterns with two- (D2), four- (D4) and six-fold (D6) symmetries in BEC with atomic interactions driven at two frequencies. To understand the underlying mechanism of non-Gauss pattern formation in complex quantum entanglement, we attempt to have an innovative approach from the perspective of complex adaptive quantum systems. Inspired by Shi's trading volume-price wave equation in finance, we define a non-localized momentum, find a non-localized quantum wave equation by the strict mathematic derivatives Schrödinger used in 1926 [30], and apply the theory to justify non-Gauss distribution in EPR experiments.

Quantum entanglement is "spooky actions at a distance" and a fantastic phenomenon. A mathematical expression defines quantum entanglement as a non-separable direct product of states [23] [24]. It is steerable and non-localized in a bipartite system in which parts A and B are entangled through interactions. It shows mutual or reciprocal causation between Parts A and B. After spatially splitting the system, measurements on the two parts yield strongly correlated outcomes, allowing one to use measures on A to predict properties of B and vice versa [26]. In a word, part A learns adaptation in response to part B in an entangled pair, and vice versa [31].

However, there has been no consensus on the interpretation of quantum entanglement. For example, what is the underlying coherent pattern formation mechanism through interactions in quantum entanglement? It has been fiercely debated for 90 years, plagued the theoretical physics community, and remains unresolved today [32] [33]. Traditionally, mainstream interpretation refers to quantum entanglement due to the superposition of coherent states in a bipartite complex quantum system. It is a paradox since quantum states with different eigenvalues are orthogonal, unitary, and independent in a Hilbert space, contradictory to entangled states. The other popular interpretation is that a quantum system is entangled through interactions since parts A and B retain the momentum and energy conservation in an entangled pair. Generally, "the Copenhagen interpretation of quantum mechanics appears in the dilemma structure: the microscopic world is described by quantum mechanics, which is unitary, but its observation or measurement depends on the classical world outside the quantum system" [33].

A probability wave measures the intensity of a wave by probability instead of amplitude. A probability wave equation is an ideal mathematical methodology in which we quantify uncertainty by a probability wave function and determine a stationary or equilibrium state by an eigenvalue. The methodology is successful in quantum mechanics and finance. For example, Schrödinger's wave equation correctly predicts the transition between energy levels in a hydrogen atom in quantum mechanics [30]. Shi's wave equation uncovers the underlying mechanism of intraday dynamic market equilibrium in finance [12]. From a complexity sciences perspective, both equations describe complex adaptive systems mathematically. However, there is a significant deviation in predicting transitions between energy levels in atoms, in which the number of atomic nuclei increases significantly. Today, physicists usually apply two methods to solve the problem: renormalization and quantum field theory [34].

Experiments in quantum mechanics reveal the quantum violation of Bell inequality [35-37]. It confirms quantum nonlocality and signifies that Newton's second law does not apply to the microscopic world in physics. It suggests redefining quantum momentum and momentum force and finding a non-localized wave equation in quantum mechanics.

We find a non-localized wave equation in quantum mechanics if cumulative observable in a time interval represents non-localized momentum or momentum force in Skinner-Shi (reinforcement-frequency-interaction) coordinates [38]. It is mathematically identical to Shi's trading volume-price probability wave equation in finance. It implies that the underlying mechanism of interactively coherent states exists in quantum mechanics and finance. Skinner-Shi (S-S) coordinates put Skinner's coordinates (reinforcement-frequency coordinates) and an interactively coherent state coordinate, Z, together. Here, the Z coordinate represents a steerable and multilevel state Hilbert space (shown in Fig. 2). It helps reveal the underlying mechanism of quantum entanglement through interactions.

Quantum entanglement as a resource is energy-consumed. Our theory shows that complex quantum entanglement follows interaction conservation instead of energy conservation over a reinforcement range. It contrasts with the widespread agreement on energy conservation in an entangled system. An interactive eigenvalue reveals a law of interaction conservation between the momentum and reversal forces over a reinforcement range. The opposite forces coexist and keep the opposite states of particles non-separable in an entangled state (shown in Fig. 1).

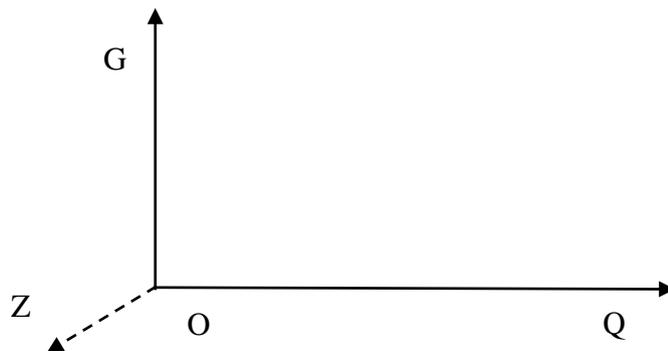

Fig. 2: Skinner-Shi or intelligent adaptive learning coordinates

Note: 1) *Q* represents reinforcement, *G* represents the operant frequency or probability, and *Z* represents steerable multilevel states or a set of discontinuous, countless, but limited states of interactive eigenvalue states in Hilbert spaces; 2) *O* is origin in the coordinates.

The paper's innovation includes 1) finding a unified paradigm for a non-localized quantum wave equation and Schrödinger's wave equation; 2) providing a testable innovative interpretation of non-Gauss entanglement distributions in complex quantum entanglement and patterns in BEC when we apply a law of interaction conservation in complex adaptive quantum systems; 3) mathematics and physics are usually applied to study economics and finance. Few explore a reverse application. Inspired by Shi's wave equation in finance, we find a non-localized wave equation in quantum mechanics.

This theoretical paper lacks experimental falsification. Fortunately, previous EPR experiments in quantum physics support it. We are seeking relevant cooperation and looking forward to seeing more experimental results.

## Methods

Schrödinger's wave equation in quantum mechanics and Shi's trading volume-price probability wave equation in finance measure uncertainty by probability wave functions and determine steerable stationary or equilibrium states by eigenvalues. From a complexity science perspective, both equations, from entirely different fields, describe the behavior of complex systems. They have commonalities and an inherent logical relationship.

A financial market is typically a complex adaptive system in which the market traders are intelligent and learning rather than ever-rational [12] [39] [40]. Shiller [41] verifies that the stock prices move too much to be justified by subsequent changes in dividends in the United States market. The market returns are about 5 to 13 times the future dividends calculated by a net present value model (NPV). Despite the excessive fluctuation, the empirical result supports that the macroeconomic cycles match the market dynamics around the market fundamentals over long horizons. The market returns behave in short-term momentum [42-44] and long-term reversals [45], reflecting the trader's adaptive intelligence. Many intelligent behaviors include thinking, communication, verbal language, perception, emotion, memory, working for a goal, computation, generation, cooperation, adaptive learning, Etc. This article focuses on intelligence capable of complex adaptive learning in response to reinforcement information in feedback loops as many units, elements, or agents interact in an open complex system. Adaptive intelligence plays a crucial role in studying, explaining, and understanding complex adaptive systems through emerging hidden patterns.

Discovering a universal law in complexity sciences through in-depth research on financial markets is an efficient and feasible research approach. For example, various characteristic correlation structure patterns exist in the observation time window, which can be classified into several typical "market states" [46]. In addition, a Chinese scientist has discovered a trading volume-price probability wave equation and gotten two sets of explicit solutions for cumulative trading volume distribution over a price range, achieving a theoretical breakthrough [9]. Its analytical solutions reveal the underlying trading mechanisms of log-normal distribution, uniform distribution, and the square of zero-order Bessel function distribution. It helps identify intraday dynamic market equilibrium states, multilevel equilibrium states, the jump of an equilibrium point between any two equilibrium levels, and holistic dynamic market disequilibrium in financial markets. It helps find a universal law in complex adaptive systems [12]. Some evidencing data from the Chinese stock markets support the proposed equation and models (see subsection Data supports and citation).

## A trading volume-price probability wave equation in finance

Shi's trading volume-price probability wave equation is written by

$$\frac{B^2}{V}\left(p\frac{d^2\psi}{dp^2} + \frac{d\psi}{dp}\right) + [E - U(p)]\psi = 0. \tag{1}$$

and

$$E(p,t) = pv_{tt} = p\frac{v_t}{t} = p\frac{v}{t^2}, \tag{2}$$

$$U(p) = A_{tt}(p - p_0), \tag{3}$$

where $p$ is a trading price (variable), $p_0$ is the market equilibrium price (a constant) at which cumulative trading volume achieves the maximal; $\psi(p)$ is a wave function, whose square $|\psi(p)|^2$ represents the probability of cumulative trading volume at the price $p$; $t$ is a time interval, usually selecting one for convenience; lowercase $v$ is a cumulative trading volume in the time interval $t$, representing operant trading frequency at the price $p$ in the time interval $t$., $v_t = \frac{v}{t} = v/t$ is a cumulative trading momentum, $v_{tt} = \frac{v}{t^2} = v/t^2$ is a cumulative trading momentum force; The capital letter $V$ represents the sum of cumulative trading volume across all trading prices in the time interval $t$, which is the total trading volume; $B$ is a dimensional constant that maintains the phase of the wave function as dimensionless; $E(p,t)$ in equation (2) is the liquidity of cumulative trading amount at the price $p$ in a time interval $t$, that is, cumulative trading energy; $U(p)$ in equation (3) is the price potential formulated by the market shortage or surplus relative to the market equilibrium price $p_0$; and $A_{tt}$ is a cumulative trading reversal force and signifies the same dimension as that of momentum force $v_{tt}$. The momentum and reversal traders are contrarian and coexist in the financial markets.

### *Data support and citation*

Equation (1) has been falsified and valid through three sets of intraday high-frequency trading data in Chinese stock markets (see the three sets of data supplements). The first data set was the high-frequency trading data of the top 30 stocks on the Shanghai 180 Index in June 2003. It supports and validates the square of zero-order Bessel functions solved from the trading volume-price probability wave equation (see Supplemental 30ShanghaiIndex) [9]. Another data set is the high-frequency trading data of Huaxia SSE (Shanghai Stock Exchange) 50ETF (510050) from April 2007 to April 2009. China's stock market experienced a complete market cycle from a bull market to a bubble market, a bubble burst market, a plunging market, and a reversal market in two years. It examines the theoretical model based on conditioning to explain investor behavior and shows that investors can be overconfident or panicked based on price momentum. The strongest positive correlation in behavior occurs during price reversals when many investors are more likely to sell their risky assets in a panic (see Supplemental HuaxiaSSE50ETF2007-2009) [47-49]. The rest of the data is collected from the Huaxia SSE 50ETF (510050) in January and February 2019. It documents that interacting trading behavior is crucial in the nonlinear dynamic price model with trading volume weights. The square of zero-order Bessel distribution disappears if the market traders are independent (see Supplemental HuaxiaSSE50ETF510050in2019) [50]. These data from the Chinese stock markets support the proposed models solved in equation (1).

These empirical results indicate that the mathematical method of the trading volume-price probability wave equation successfully models dynamic market behaviors. The probability wave eigenfunctions with eigenvalues solved from the wave equation capture uncertain, nonlinear, and non-monotone price fluctuation and asset prices in complex financial markets. The proposed models reveal the underlying mechanism of local dynamic market equilibrium, multilevel equilibrium, and holistic market disequilibrium. Momentum traders drive the market prices to diverge from an equilibrium point, and reversal traders return the prices. Momentum and reversal traders coexist. Their contrarian behaviors are a kind of self-organization in complex adaptive systems [51]. Interactive traders keep interactive coherence. The market emerges a nonlinear and non-monotonic intraday cumulative trading volume distribution over a price range (see Fig. 3). The jump in equilibrium prices results in multilevel equilibrium and holistic dynamic market disequilibrium since partly equilibrium states have different equilibrium points. Intelligent traders are adaptive and learning instead of ever-rational (see Fig. 4) [12].

Automated Trader, a financial industry magazine in the United Kingdom, reported the research in 2013 [48]. In 2022, a research team led by Professor Andrew W. Lo from the Sloan School of Management at the Massachusetts Institute of Technology in the United States cited a 2011 working paper [49].

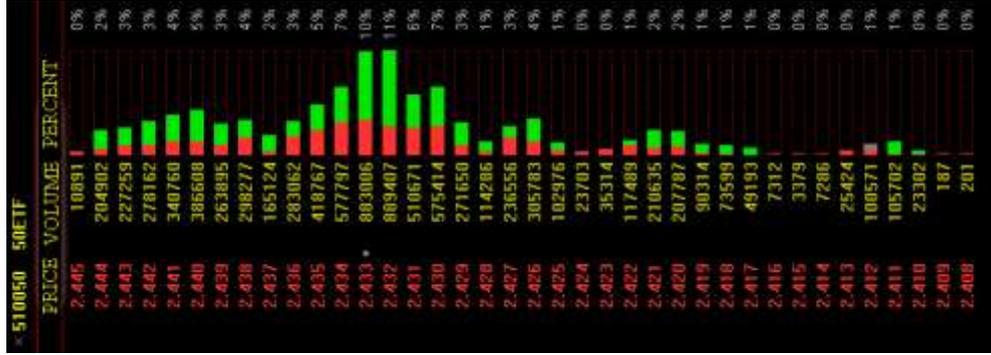

Fig. 3: A price list with intraday cumulative trading volume (raw-data screenshot).

Note 1) The data collected from Huaxia SSE 50ETF (510050) dated January 25, 2019, in the Shanghai Securities Exchange in the Chinese stock market; 2) When the momentum traders drive the price to diverge, the reversal traders will bring the price back to the equilibrium price, at which the corresponding cumulative trading volume achieves the maximum value in intraday dynamic market equilibrium. 3) The volume unit is 100 shares.

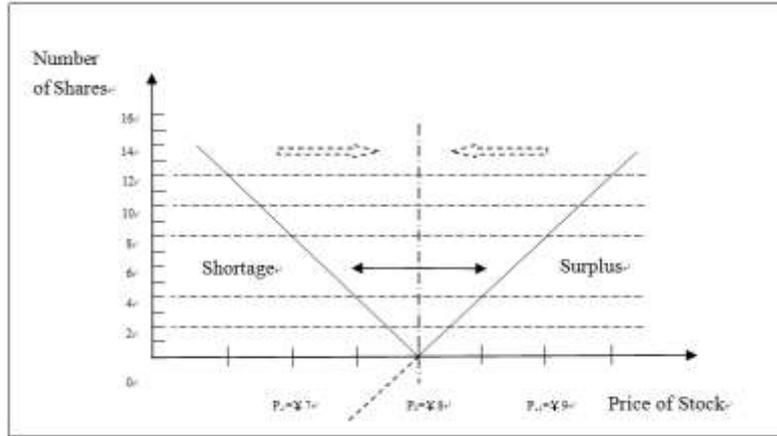

Fig. 4: A linear V-shaped shortage and surplus curve illustrates intraday dynamic market equilibrium

Note 1) The simplified solid V-shaped curve is a shortage and surplus curve; 2) $P_0$ is an equilibrium price in intraday dynamic market equilibrium where the shortage and the surplus are zero; 3) The negative surplus (the dotted surplus line) represents the shortage in a linear surplus line; 4) The double-headed solid arrow represents the direction of momentum trading; 5) The dashed left and right arrows signify the direction of reversal trading; 6) ¥ represents Chinese currency RMB yuan.

**A law in complex financial markets**

We extract a law of invariance of interaction from a trading volume-price probability wave equation (1) in complex adaptive financial markets [12]. It is expressed mathematically by

$$\omega_n^2 = \frac{v_{t,n,i}^2}{V} = \frac{v_{n,i}}{V} v_{tt,n,i} = v_{tt,n,i} - A_{tt,n,i} = const. \qquad (n = 0,1,\cdots), (i = 0,1,2\ldots) \qquad (4)$$

where $v_{n,i}$ is a cumulative trading volume which can represent operant trading frequency by $v_{n,i}/V$ at a price $p_i$, $v_{t,n,i}$ is a cumulative trading momentum, $v_{tt,n,i}$ is a cumulative trading momentum force, $A_{tt,n,i}$ is a cumulative trading reversal force; $\omega_n$ or $\frac{v_{n,i}}{V} v_{tt,n,i}$ is a cumulative trading interaction frequency or interaction force between momentum force $v_{tt,n,i}$ and reversal force $A_{tt,n,i}$; $n$ represents the $n^{th}$ state, and $i$ is the $i^{th}$ point in a sampling size.

Equation (4) specifies interaction conservation over a price range. It is a mathematical expression for game equilibrium in a dynamic market economy. Next, we will find a non-localized wave equation and a law in quantum mechanics.

**A non-localized wave equation in quantum mechanics**

Experimental tests violate Bell's inequality in quantum mechanics. It confirms a quantum non-localized property [35-37]. The momentum and momentum force does not depend on distance, displacement, velocity, or acceleration, and Newton's second law does not apply to the microscopic world in quantum mechanics. It says we must redefine momentum and momentum force in complex adaptive quantum systems and find a non-localized wave equation. Quantum mechanics supports this idea. We measure particle uncertainty by wave function, determine stationary states by eigenvalues, and predict transition between energy levels by energy conservation in quantum mechanics.

Inspired by a trading volume-price probability wave equation in the financial market, we have three assumptions about non-localized variables as follows:

**Assumption Ⅰ:** The non-localized momentum $Q$ is a cumulative observable $m$ at a reinforcement point $q$ in a time interval $t$. It is defined by equation (5).

**Assumption Ⅱ:** The non-localized momentum force $F$ is the momentum $Q$ in the time interval $t$. It is defined by equation (6).

**Assumption Ⅲ:** The energy is the product of non-localized momentum force $F$ and a generalized reinforcement $q$. It is defined by equation (7).

Let
$$Q \equiv \frac{\partial S(q,t)}{\partial q} = \frac{m}{t} = m_t, \tag{5}$$

$$F \equiv \frac{Q(q,t)}{t} = \frac{m_t}{t} = \frac{m}{t^2} = m_{tt}, \tag{6}$$

$$E(q,t) = F * q = qm_{tt} = q\frac{m}{t^2} = q\frac{m_t}{t}, \tag{7}$$

where $S(q,t)$ is action; $q$ is a generalized reinforcement coordinate and can be an electronic field, a magnetic field, or a light field to which particles are sensitive; $m$ is a cumulative observable at a point $q$ in a time interval $t$; $t$ is a time interval, usually choosing one for convenience; $m_t$ is non-localized momentum $Q$ and the cumulative observable $m$ over the time interval $t$; $m_{tt}$ is the non-localized momentum force $F$ in the reinforcement-frequency-eigenvalue coordinates; $E(q,t)$ is the product of non-localized force $F$ and reinforcement coordinate $q$, representing particle's cumulative energy at a point $q$ in the reinforcement coordinates. The momentum $Q$ and force $F$ depend on cumulative observable $m$ in a time interval $t$.

An identical equation (8) holds in complex adaptive quantum systems. It is an interdependent causal rule, and Soros calls the reciprocal or mutual causation reflexivity [52]. Thus, energy holds in a rule as follows,
$$E(q,t) \equiv PE(q,t) + (1-P)E(q,t) = PE(q,t) + U(q-q_0), \tag{8}$$
and
$$P = \frac{m}{M}, \tag{9}$$

where $q$ is a non-localized coordinate or reinforcement coordinate, $q_0$ is a singular or equilibrium point; $m$ is a cumulative observable at a point $q$ in a time interval $t$, $M$ is a total cumulative observable over a range of reinforcement in the time interval $t$; $P$ is a probability at $q$; $E(q,t)$ is particle's cumulative energy at a point $q$ in the reinforcement coordinate, $PE(q,t)$ is interaction or distribution energy in the time interval, and $U(q)$ is reversal energy or potential in the non-localized complex adaptive quantum systems.

Putting equation (7) and (9) into the right side of equations (8), we have
$$E(q,t) = PE(q,t) + U(q-q_0) = \frac{m}{M}\left(q\frac{m}{t^2}\right) + U(q-q_0) = \frac{1}{M}q\left(\frac{m}{t}\right)^2 + U(q-q_0), \tag{10}$$

then, we can rewrite equation (10) as

$$-E(q,t) + q\frac{m_t^2}{M} + U(q - q_0) = 0. \tag{11}$$

If reversal or potential energy is written as

$$U(p) = A_{tt}(p - p_0), \tag{12}$$

where $A_{tt}$ is the reversal force in the linear potential $U(q)$. Differential two sides of equation (11) by $q$, we have

$$-m_{tt} + \frac{m}{M}m_{tt} + A_{tt} = 0. \tag{13}$$

Assume that an unknown function $\psi(q,t)$ is as follows [30] [53] [54]

$$\psi(q,t) = Re^{iS(q,t)/B}, \tag{14}$$

where $q$ is a reinforcement coordinate, $t$ is a time interval, $R$ is amplitude; the function $S(q,t)$ is action, assumed to be twice differentiable in all its arguments, and known as Hamilton's principal function; $B$ is a dimensional constant to keep phase $S(q,t)/B$ dimensionless, $i$ is an imaginary number and $i^2 = -1$, and $\psi(q,t)$ is an unknown wave function to be determined.

From equation (14), we have

$$\frac{\partial S(q,t)}{\partial q} = -\frac{iB}{\psi(q,t)}\frac{\partial \psi(q,t)}{\partial q}. \tag{15}$$

Assume that particles abide by the Hamilton-Jacobi equation in non-localized coordinates [55] as

$$\frac{\partial S}{\partial t} + H\left(q, \frac{\partial S}{\partial q}\right) = 0, \tag{16}$$

where $q$ is a reinforcement coordinate, $S(q,t)$ is action, $t$ is a time interval, and $H\left(q, \frac{\partial S}{\partial q}\right)$ is Hamiltonian.

By equation (11), we define Hamiltonian as

$$H\left(q, \frac{\partial S}{\partial q}\right) = q\frac{m_t^2}{M} + U(q - q_0). \tag{17}$$

Equation (16) is satisfied in energy conservative systems. However, we can apply it to complex dissipative quantum systems for three reasons: 1) equation (10) or (11) is based on an identical equation (8) and holds no matter whether the systems are energy-conservative; 2) if $q_0 \gg q - q_0$ and $M \gg m$, then $E$ or $U \gg p\frac{v_t^2}{V}$. In this case, $E$ can be regarded as infinity and close to a constant; 3) if the non-localized wave equation obtained in the trial and error process can correctly describe the particle's behaviors and reveal underlying mechanisms of quantum entanglement, then the assumption equation (16) is acceptable.

By separating variables, we rewrite equation (16) as,

$$H\left(q, \frac{\partial S}{\partial q}\right) = -\frac{\partial S}{\partial t} = E, \tag{18}$$

then, we have two equations using equation (17) as follows

$$-\frac{\partial S}{\partial t} = E, \tag{19}$$

and

$$q\frac{m_t^2}{M} + U(q - q_0) = E. \tag{20}$$

From equation (19), we have

$$S(q,t) = S_1(q) - Et. \tag{21}$$

In equation (21), $S_1(q)$, $Et$, and $qm_t$ have the same dimension. We can have a special solution as

$$S(q,t) = \alpha(qm_t) - Et + \beta \equiv qm_t - Et, \tag{22}$$

where $\alpha$ or $\beta$ is any constant. We choose one for convenience.

Thus, we have the momentum $Q$ and momentum force $F$ from equation (22) as follows

$$Q \equiv \frac{\partial S}{\partial q} = m_t, \tag{23}$$

and

$$F \equiv \frac{Q}{t} = m_{tt}. \tag{24}$$

Equations (23) and (24) are assumptions Ⅰ and Ⅱ, respectively.

Replaced by equation (23), equation (20) can be written by

$$-E + \frac{q}{M}\left(\frac{\partial S}{\partial q}\right)^2 + U(q - q_0) = 0. \tag{25}$$

From equations (15) and (25), we can construct a Lagrange functional $L(q,\psi)$ using the properties of conjugative functions as follows,

$$L(q,\psi) \equiv \varepsilon = (U - E)\psi^*\psi + \frac{B^2}{M}p\left(\frac{\partial \psi^*}{\partial q}\right)\left(\frac{\partial \psi}{\partial q}\right). \tag{26}$$

Our variation problem then reads

$$\delta \int L(q,\psi)dq = 0. \tag{27}$$

By the Euler-Lagrange equation (to $\psi^*$), we have a time-independent wave equation in non-localized quantum mechanics in which the momentum and force depend on density, probability, or cumulative observable in complex adaptive quantum systems as

$$\frac{B^2}{M}\left(q\frac{d^2\psi}{dq^2} + \frac{d\psi}{dq}\right) + [E - U(q - q_0)]\psi = 0. \tag{28}$$

The non-localized wave equation (28) in quantum mechanics is mathematically identical to Shi's trading volume-price probability wave equation (1) in finance [9]. $|\psi(q)|^2$ represents observable probability in the wave equation (28). It satisfies a normalized condition as

$$\int |\psi(p)|^2 \, dp = 1 \text{ or } \sum_i |\psi_i(p_i)|^2 = 1 \tag{29}$$

**Two sets of explicit solutions from the non-localized wave equation in quantum mechanics**

Let us solve the non-localized wave equation in quantum mechanics as follows

$$\frac{B^2}{M}\left(q\frac{d^2\psi}{dq^2} + \frac{d\psi}{dq}\right) + [E - U(q - q_0)]\psi = 0, \tag{30}$$

subject to natural boundary conditions in a non-localized and one-dimensional reinforcement coordinate

$$\begin{cases} \psi(+\infty) & \to 0 \\ \psi(q_0) & < \infty. \\ \psi(-\infty) & \to 0 \end{cases} \tag{31}$$

From equations (30) and (31), we get two sets of explicit solutions in the non-localized coordinates. Its mathematical expressions are identical to those in Shi's trading wave equation [9, 12].

One is a set of multi-order eigenfunctions with reversal force eigenvalue. The non-localized force or energy $E$ is a constant over a reinforcement range. It is an energy-conservative system. In this case, particles are independent because there is no interaction in eigenvalues.

The square of eigenfunctions is a set of log-normal or uniform wave functions. The wave functions are written by

$$\psi_{n,i}(q_i) = C_n e^{-\sqrt{A_{tt,n}}|q_i - q_0|} \cdot F\left(-n, 1, 2\sqrt{A_{tt,n}}|q_i - q_0|\right), \quad (n, i = 0,1,2\cdots) \tag{32}$$

or

$$|\psi_{n,i}(q_i)|^2 = C_n e^{-2\sqrt{A_{tt,n}}|q_i - q_0|} \cdot \left|F\left(-n, 1, 2\sqrt{A_{tt,n}}|q_i - q_0|\right)\right|^2, \quad (n, i = 0,1,2\cdots) \tag{33}$$

subject to reversal force eigenvalues in a linear potential $U(q_i)$ as follows

$$\sqrt{A_{tt,n}} = \frac{E_n}{1+2n} = constant > 0, \quad (n = 0,1,2\cdots) \tag{34}$$

or

$$E_n = (1 + 2n)\sqrt{A_{tt,n}} = (1 + 2n)\omega_n = constant > 0, \quad (n = 0,1,2\cdots) \tag{35}$$

where $|\psi_{n,i}(q_i)|^2$ is the particle's cumulative observable probability at a point $q_i$ in a conservative system where energy $E_n$ is a constant, expressed by equation (35), over a reinforcement range; $-A_{tt,n}$ is a reversal force eigenvalue (a constant), and the minus sign means that the direction of the reversal force is always to a singular point $q_0$; $\omega_n^2 = A_n$ ($\omega_n > 0$) is the frequency of the wave function, equation (33); $U(q_i)$ is a V-shaped linear potential function in equation (12); $C_n$ is a normalized constant, $n = 0,1,2,3\cdots$ is the order in multi-order eigenfunctions, and $i = 0,1,2,3\cdots$ represents the $i^{th}$ point in a sampling size, $\sqrt{A_n}$ is reversal force eigenvalues, $E_n$ is energy; $F(-n, 1, 2\sqrt{A_n}|q_i - q_0|)$ is a set of $n$-order confluent hypergeometric eigenfunctions (the first Kummer's eigenfunctions) $F(\alpha, \gamma, \xi)$. It is defined as

$$F(\alpha, \gamma, \xi) = 1 + \frac{\alpha}{\gamma}\xi + \frac{\alpha(\alpha+1)}{2!\gamma(\gamma+1)}\xi^2 + \frac{\alpha(\alpha+1)(\alpha+2)}{3!\gamma(\gamma+1)(\gamma+2)}\xi^3 + \cdots$$
$$= \sum_{k=0}^{\infty} \frac{(\alpha)_k}{k!(\gamma)_k}\xi^k, \tag{36}$$

and  $(\alpha)_k = \alpha(\alpha+1)\cdots(\alpha+k-1),$ (37)
$(\gamma)_k = \gamma(\gamma+1)\cdots(\gamma+k-1),$ (38)

where $F(\alpha, \gamma, \xi)$ has meaning if and only if $\gamma \neq 0$ and $\alpha = -n$ ($n = 0,1,2\cdots$) are satisfied.

$|\psi_{n,i}(q_i)|^2$ in equation (33) is plotted in Fig. 5 if $n$ is 0, 1, 2, and 10, respectively. It is an independent or conservative system since the eigenvalues do not have interaction information, and energy is a constant expressed by equation (35).

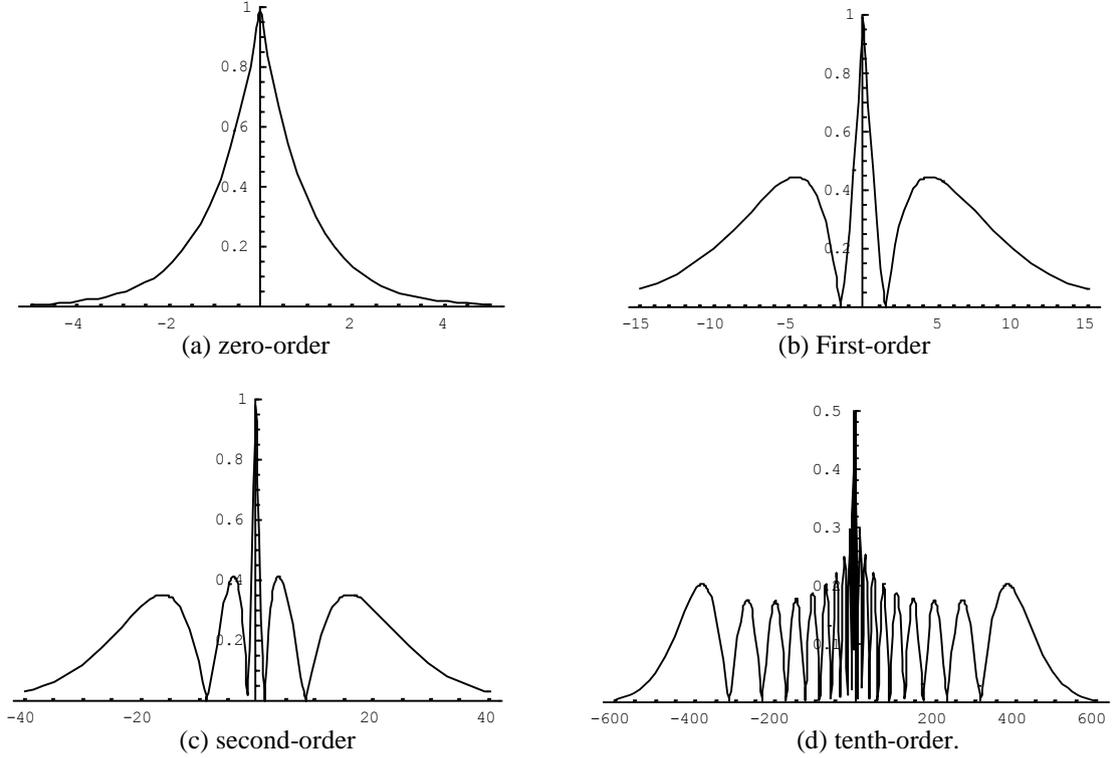

(a) zero-order  (b) First-order
(c) second-order  (d) tenth-order.

Fig. 5: Multi-order eigenfunctions in independent trading

Note: The horizontal coordinate represents reinforcement, the vertical coordinate represents cumulative observable, and the origin is a singular point in equation (28) or (30).

The other is a set of zero-order Bessel eigenfunctions if cumulative energy $E_i = q_i m_{tt,i}$ is a separable variable. A particle's probability or distribution over a reinforcement range obeys the square of a set of zero-order Bessel eigenfunctions. It is written by

$$\psi_{n,i}(q_i) = C_n J_{0,i}[\omega_n(q_i - q_0)], \quad (n = 0,1\cdots), (i = 1,2\ldots) \tag{39}$$

or

$$|\psi_{n,i}(q_i)|^2 = C_n |J_{0,i}[\omega_n(q_i - q_0)]|^2, \quad (n = 0,1\cdots), (i = 1,2\ldots) \tag{40}$$

where $q_i$ is a reinforcement coordinate, $q_0$ is a singular point or the level of the stationary state, $\omega_n$ is an interactive eigenvalue or frequency, $J_0$ is a set of zero-order Bessel eigenfunctions, $|\psi(q_i)|^2$ represents cumulative observable probability at $q_i$; $C_n$ is a normalized constant, $n$ represents the $n^{th}$ state, and $i$ is the $i^{th}$ point in a sampling size (see Fig. 6);. The reinforcement can be any a thing, to which particles are sensitive, such a leaser field, an electric field, or a magnetic field.

Equation (39) or (40) holds if and only if the interactive eigenvalues or frequencies $\omega_n$ satisfy

$$\omega_n^2 = \frac{m_{t,n,i}^2}{M} = \frac{m_{n,i}}{M} m_{tt,n,i} = m_{tt,n,i} - A_{tt,n,i} = const. \qquad (n = 0,1,\cdots), (i = 0,1,2\ldots) \qquad (41)$$

where $m_{tt,n,i}$ is a particle's cumulative momentum force, $A_{tt,n,i}$ is the cumulative reversal force and $\frac{m}{M}m_{tt,n,i}$ is non-separable interactive force; $n$ represents the $n^{th}$ state, and $i$ is the $i^{th}$ point in a sampling size.

Equation (41) is mathematically identical to equation (4).

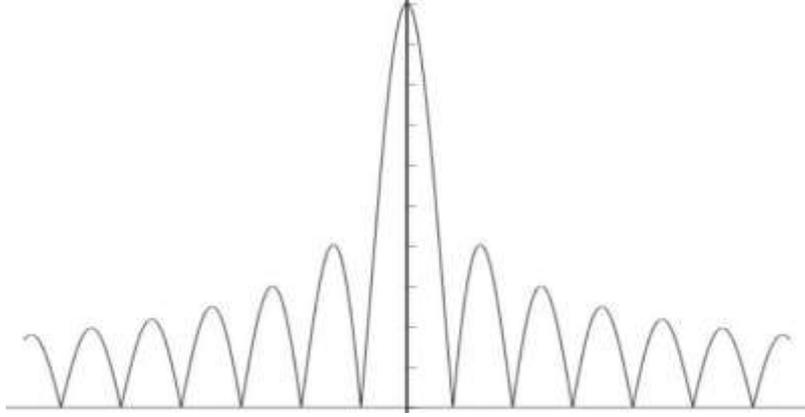

Fig. 6: The square of zero-order Bessel distribution over a reinforcement range

Note: The horizontal coordinate represents reinforcement, and the vertical coordinate represents cumulative observable probability in a time interval. The origin is a singular point.

**Schrödinger's wave equation and a non-localized wave equation in quantum mechanics**

Schrödinger finds a particle's wave equation in terms of a principle of energy conservation in classical mechanics. It accurately predicts the transition of an electron between energy levels in a hydrogen atom (non-relativistic and unperturbed) [30]. We will find Schrödinger's wave equation from a complex sciences perspective.

*Schrödinger's wave equation from complex systems*
Assume that

$$E \equiv PE + (1-P)E = D(x) + U(x), \qquad (42)$$

where $E$ is energy, which equals the work expressed by the product of force $f$ and distance $x$; $P$ is the observable probability of particles at a position $x$; $D(x)$ represents distribution energy, which is the product of the probability $P$ and energy $E$; the rest is potential energy $U(x)$.

Equation (42) is an identical equation and holds in open complex systems. If the momentum force $f$ obeys Newton's second law, $f = ma = m\frac{x}{t^2}$, as Schrödinger assumed in his seminal work, then equation (42) is written as follows,

$$-E + \frac{p^2}{M} + U(x) = 0, \qquad (43)$$

and

$$D(x) = PE = \frac{m}{M}E = \frac{m}{M}f * x = \frac{m}{M}(ma) * x = \frac{m^2}{M}\left(\frac{x}{t}\right)^2 = \frac{p^2}{M}, \qquad (44)$$

where $p = m\frac{x}{t}$ is momentum, $m$ is cumulative observable at the position $x$, and $M$ is total observables in the complex systems.

From equation (43), we can obtain Schrödinger's wave equation in open complex systems. It is as follows.

$$\left[-\frac{\hbar^2}{M}\frac{d^2}{dx^2} + U(x)\right]\psi(x) = E\psi(x). \qquad (45)$$

*Schrödinger's wave equation and a non-localized wave equation*

Equation (45) is equivalent to Schrödinger's wave equation in quantum mechanism. Both are based on the assumption that the momentum force obeys Newton's second law.

Schrödinger's wave equation correctly predicts the transition between energy levels in the hydrogen atoms. However, there is a significant deviation in predicting transitions between energy levels in atoms, where the number of atomic nuclei is more significant when using this equation. Currently, two main methods are used to solve the deviation problem: renormalization and quantum field theory [34].

Fig. 7 illustrates an inherent logical relationship between Schrödinger's wave equation and the non-localized wave equation in quantum mechanics.

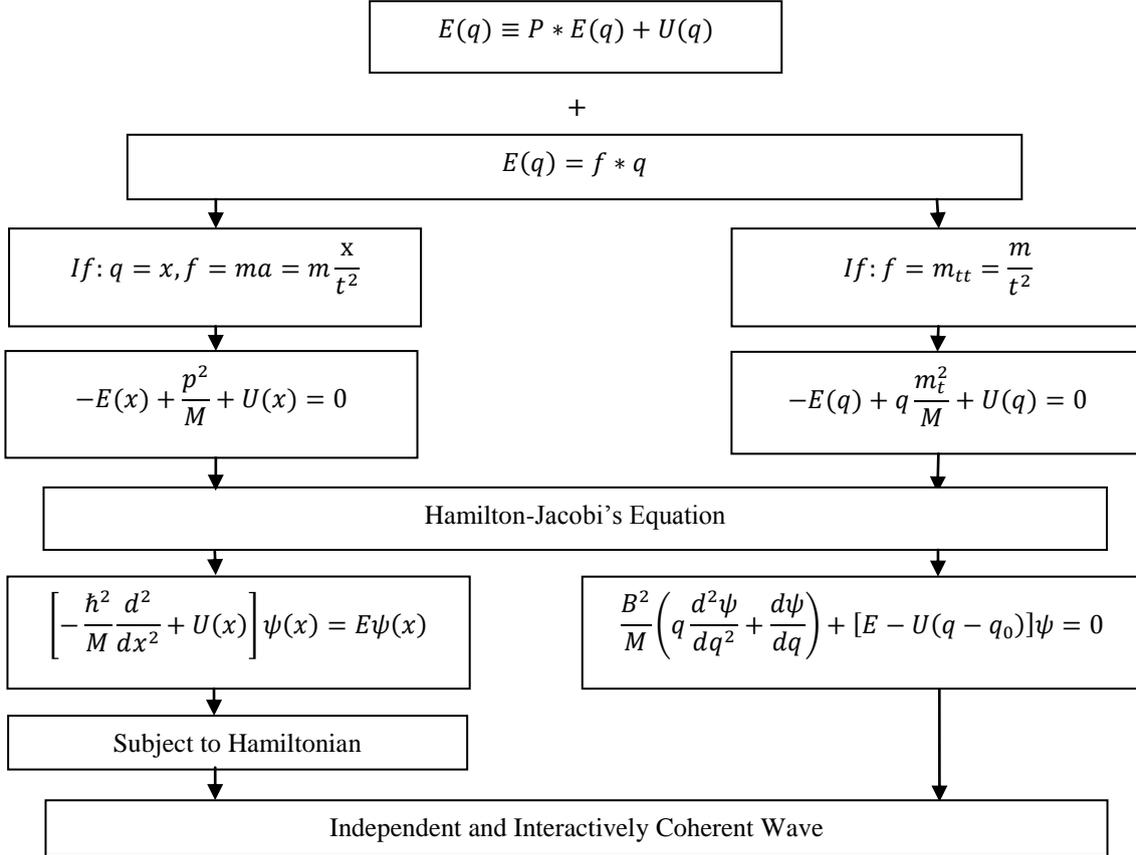

Fig. 7: A reciprocally or mutually causal paradigm

Note: 1) Energy $E$ represents the work that equals the product of momentum force $f$ and generalized distance $q$, and $P$ represents observable probability which is equal to $m/M$ at a $q$ point; 2) Schrödinger's wave equation governs the probability wave if momentum force obeys Newton's law (see left column); 3) a non-localized wave equation, equation (28) or (30), holds if momentum force is cumulative observable in a time interval. The momentum force is non-localized and probability-dependent, violating Newton's second law (see right column).

## Results

Experiment results in physics violate Bell inequality. It confirms that quantum is non-localized regardless of its position, displacement, velocity, and acceleration. Thus, we redefine momentum and momentum force in complex adaptive quantum systems and find a non-localized wave equation (28) or (30) in quantum mechanics.

### Two-worlds in quantum mechanics

We propose two-world interpretations in quantum mechanics. One is based on Schrödinger's wave equation in classical quantum mechanics. The other is based on a non-localized wave equation (28) or (30) (see Fig. 7). The two-world interpretation can be traced back to Everett's many-world interpretations (MWI) [33] [56]. We summarize the characteristics of the two worlds shown in Table 1.

| Characteristics | Measurement | Schrödinger's wave equation | Shi's non-localized wave equation |
|---|---|---|---|
| Independent particles or agents | No interaction (disorder) | Yes | Yes |
| Interacting particles or agents | Hidden patterns | Subject to Hamiltonian | Yes |
| Uncertainty | Probability wave function | Yes | Yes |
| Stationary states | Eigenvalue | Yes<br>Energy eigenvalue | Yes<br>Interactive eigenvalue |
| Transition | Energy or reference point levels | Yes<br>Energy Levels | Yes<br>Reference point levels |
| Newton's laws | $f=m\,x/t^2$ | Yes | No |
| Locality | Distance $x$ | Yes | No |
| Non-localized force | $f=m/t^2$ | No | Yes |
| Energy conservation | $E=K+U$ | Yes | Yes<br>Equation (35) |
| An interdependent energy rule | $E=PE+U$ | Yes<br>Equation (42) | Yes<br>Equation (8) |
| Interaction conservation | Equation (41) or (48) | No | Yes |
| Two worlds | Coordinates | Newton-Schrödinger (classical quantum mechanics) coordinates | Skinner-Shi (reinforcement-frequency-interaction) coordinates |

Table 1: Schrödinger's wave equation and Shi's non-localized wave equation in quantum mechanics

Note: 1) $F=ma=m(x/t^2)$ is Newton's second law in classical dynamics; 2) Cumulative observable in a time interval $m/t^2$ represents a non-localized momentum force in complex adaptive quantum systems; 3) $E$ is energy, $K$ signifies kinetic energy, $U$ denotes potential energy, and $PE$ represents distribution or interaction energy, respectively.

## A law in complex financial markets, complex adaptive quantum systems, and complex adaptive systems

An invariance of interaction exists between a momentum trading force $v_{tt}$ and a reversal trading force $A_{tt}$ in complex financial markets. It is expressed by equation (4). In addition, an invariance of interaction exists between a momentum force $m_{tt}$ and a reversal momentum force $A_{tt}$ in non-localized complex quantum systems. It is written by equation (41).

From two examples, we further assume that invariance of interaction is applicable in complex adaptive systems cutting across disciplines from social to natural sciences. It is universal and written by

$$\omega_n^2 = \frac{m_{t,n,i}^2}{M} = \frac{m_{n,i}}{M} m_{tt,n,i} = m_{tt,n,i} - A_{tt,n,i} = const. \qquad (n = 0,1,\cdots), (i = 0,1,2\ldots) \qquad (46)$$

Equation (46) is a law of interaction conservation over a reinforcement range in complex adaptive systems.

## An innovative interpretation of non-localized quantum entanglement

Quantum entanglement is non-localized in complex adaptive quantum systems. It is distributed over a reinforcement range to which entangled particles are sensitive. The distribution is written as follows

$$|\psi_{n,i}(q_i)|^2 = C_n |J_{0,i}[\omega_n(q_i - q_0)]|^2, \qquad (n = 0,1\cdots), (i = 1,2\ldots) \qquad (47)$$

subject to

$$\omega_n^2 = \frac{m_{t,n,i}^2}{M} = \frac{m_{n,i}}{M} m_{tt,n,i} = m_{tt,n,i} - A_{tt,n,i} = const, \qquad (n = 0,1,\cdots), (i = 0,1,2\ldots) \qquad (48)$$

where $q_i$ is a generalized reinforcement coordinate, $q_0$ is a singular point or the level of a stationary state in a non-localized wave equation (28) or (30); $\psi(q_i)$ is a wave function, whose square $|\psi(q_i)|^2$ represents particle's probability at the point $q_i$ in a time interval $t$; lowercase $m_{n,i}$ is a particle's cumulative observable in the time interval $t$ at the $q_i$ in the time interval $t$, $m_{t,i} = m_i/t$ is the particle's cumulative momentum, $m_{tt,i} = m_i/t^2$ is the particle's cumulative momentum force; $A_{tt,n,i}$ is reversal force; The capital letter $M$ represents the sum of the particle's cumulative observable over a reinforcement range in the time interval $t$, which is total observable; $B$ is a dimensional constant that maintains the phase of the wave function as dimensionless; $\omega_n^2$ is interaction frequency or interaction eigenvalue; a cross item $\frac{m}{M} m_{tt,n,i}$ is non-separable interaction force between momentum force $m_{tt,n,i}$ and reversal force $A_{tt,n,i}$; $n$ represents the $n^{th}$ state and $i$ is the $i^{th}$ point in a sampling size; $J_0$ is a set of zero-order Bessel eigenfunctions with interaction eigenvalues $\omega_n^2 = \frac{m_{n,i}}{M} m_{tt,n,i}$.

Equation (48) is a necessary and sufficient condition for equation (47). Equation (48) has two meanings in quantum mechanics. First, it signifies interaction conservation between momentum force $m_{tt}$ and reversal force $A_{tt}$, and the interaction frequency is a constant over a reinforcement range. Any change in one force is compensated for by an equal but opposite change in the other in quantum entanglement. Second, momentum force $m_{tt}$ and reversal force $A_{tt}$ coexist in interaction, and they entangle opposite particle's states non-separable in an entangled state, such as spin up and spin down, black and white, cat's alive and dead, Etc. Equation (46) is a wave function in an interactively coherent or entangled state.

Consequently, part A particles learn to adapt and change in response to part B particles in an interactively coherent state, and vice versa. Quantum entanglement is an interactively coherent state between opposite and adaptive forces rather than a consequence of the supposition state of coherent states.

**Data supports, measurement, and a criterion for complex quantum entanglement**

Many evidences support the non-Gauss distribution written by equation (47) shown in Fig. 6, such as non-Gauss distributions in complex quantum entanglement (shown in Fig. 8), a multiparticle entanglement distribution in BEC, and an observation pattern in the interference of BEC [25-27]. An experimental method is a key for measurement [57].

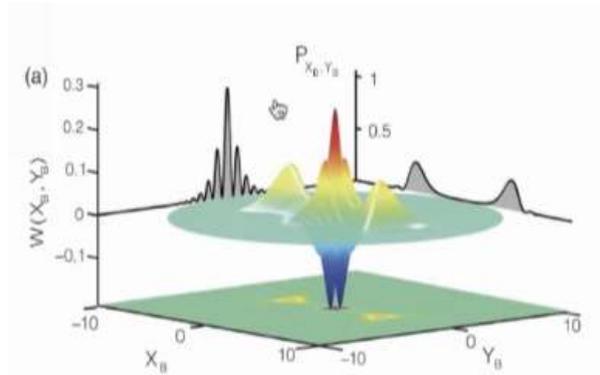

Fig. 8: Non-Gauss distribution prepared by non-local quantum steering [25]

Equation (47) is a set of explicit solutions from a non-localized quantum wave equation (28) or (30). It is a wave function in an entangled state subject to interaction conservation over a reinforcement range, equation (48). Equation (48) uncovers the underlying mechanism of quantum entanglement.

Thus, we propose equation (47) as a testable criterion for complex quantum entanglement and apply it to non-Gauss distributions in complex quantum entanglement.

## Discussion

This section will respond to some questions from reviewers, readers, and conference participants, discuss limitations, and explore the future research direction.

A. Skeptical readers argue that "the similarities between Schrödinger's wave equation and Shi's wave equation in mathematics do not support the complex adaptive learning theory, extracted from a trading volume-price probability wave equation, and could not be applied in the complex quantum systems and justify the mechanism of quantum entanglement. The authors cannot simply consider quantum particles to possess intelligence-like properties in complex adaptive quantum systems. Moreover, a typical quantum feature of a quantum particle is the quantization of energy, and the minimum energy cannot be divided. Such quantum effects can be shown through two- and high-order quantum correlation measurements, which are typically different from classical waves".

   It is a good argument. We improve our writing in three aspects as follows:

   1) Experimental tests in quantum mechanics violate Bell's inequality [32] [35-37]. It confirms a non-localized property in quantum mechanics regardless of particles' position, displacement, velocity, and acceleration. Thus, we must redefine the momentum and momentum force in complex adaptive quantum systems and find a non-localized wave equation to capture a holistic picture of particles in quantum mechanics. We find that a non-localized wave equation in quantum mechanics is mathematically identical to finance's trading volume-price probability wave equation. It suggests the exact underlying mechanism in the two fields. The study supports the assumption that invariance of interaction as a universal law exists in quantum mechanics and finance. The interaction conservation is written mathematically by equations (4) and (48).

   2) The wave-particle duality is a crucial feature of quantum mechanics. Whereas duality is a critical link between classical and quantum mechanics, it does not reject a particle's "intelligence-like" property in complex adaptive quantum systems. Entangled particles can have the third property, such as an "intelligence-like" property, although the wave-particle duality is well-known in physics. Many nonlife matters in physics show intelligent adaptive behavior in complex systems, such as self-organized criticality in a desert [58], water particles in a weather pattern, iron atoms in spin class or amorphous alloy [4], a neural network of beam-like components in a so-called architected inanimate material [59], and bipartite photons in quantum entanglement [26]. Thus, we can assume that entangled particles are "intelligence-like" in S-S (Skinner-Shi) coordinates (shown in Fig. 2).

   3) We find a unified framework for Schrödinger's wave equation and Shi's wave equation (shown in Fig. 7) and have an innovative two-world interpretation of quantum mechanics. While Schrödinger's wave equation governs one world in spatial coordinates, the non-localized wave equation works in the other world (Skinner-Shi coordinates). The non-localized wave equation reveals quantum behaviors subject to energy conservation (independent particles in a closed system) and interaction conservation (interacting particles in an open complex system), respectively (shown in Table 1).

B. Energy conservation and interaction conservation in the non-localized quantum wave equation

   The wave functions are a set of eigenfunctions, equations (33) if the energy conservation equation (35) is satisfied. If the interaction conservation equation (48) is satisfied, the wave functions are a set of eigenfunctions, equations (47).

C. A universal law beyond finance and quantum mechanics

   If the invariance of interaction is a universal law in complex adaptive systems, the law holds beyond the scope of finance and quantum mechanics. Using data from other disciplines, such as intelligence science, we can examine more distributions by the square of zero-order Bessel distribution over a reinforcement range.

D. Potential application

   The study may guide the direction of large-scale and industrialized production of quantum entangled states and supply highly demanded entanglement resources for quantum computing and quantum communications in the future.

E. Limitation

Although the theoretical work is supported by data from previous EPR experiments in quantum physics, we seek relevant cooperation and look forward to seeing more experimental results.

## Summaries and Conclusions

Quantum mechanics and finance are independent natural and social sciences disciplines, seemingly far apart. However, both successfully study the behaviors of interacting particles or traders through hidden patterns using the probability wave function in a wave equation. From a complexity sciences perspective, they have commonalities.

Quantum experiments reveal the quantum violation of Bell's inequality. It confirms a quantum non-localized property. Inspired by a trading volume-price wave equation in complex financial markets, we define non-localized momentum or momentum forces, find a non-localized wave equation in quantum mechanics, and provide a testable interpretation of complex quantum entanglement on non-Gauss entanglement distribution when we apply a law of interaction conservation written by equation (48). We can propose equation (47) as a non-localized complex quantum entanglement criterion.

We find a unified new paradigm for Schrödinger's wave equation and Shi's wave equation (shown in Fig. 7) and have an innovative two-world interpretation of quantum mechanics. We infer that entangled particles with opposite states learn to adapt, abide by a law of interaction conservation instead of energy conservation, and show an adaptive "intelligence-like" property in complex adaptive quantum systems. We conclude that quantum entanglement is an interactively coherent state between the momentum and reversal forces rather than a consequence of the superposition of coherent states. Its wave function is written by equation (47) subject to interaction conservation, equation (48). The momentum and reversal forces coexist and keep entangled particles in opposite states, such as spin up and spin down, white and black, cat's living and dead, Etc. An entangled state is non-separable and reciprocally causal in bipartite entanglement systems.

The study provides the references for experimental physicists to design new technical routes, prepare quantum entanglement resources, and test the theory's validity. It will steer the large-scale and industrialized production of quantum entangled states for highly demanded entanglement resources in quantum communications and quantum computing in the future.

## References

1. Prigogine, Ilya (1977): "Time, Structure and Fluctuations," *Nobel Lecture, Chemistry 1971-1980 (Editor-in-Charge Frängsmyr, T., Editor Forsén, S.)*, Singapore: World Scientific Publishing Co., 1993.

2. Blobel, Günter (1999): "Protein Targeting," *Nobel Lectures, Physiology or Medicine 1996-2000 (Editor Hans Jörnvall)*, Singapore: World Scientific Publishing Co., 2003.

3. Holland, John H. (1995): *Hidden Order: How Adaptation Builds Complexity*, Reading, Massachusetts; Menlo Park, California; New York: Addison-Wesley.

4. Popular information (2021): "They Found Hidden Patterns in the Climate and in Other Complex Phenomena," *The Nobel Prize in Physics 2021*, NobelPrize.org. Nobel Prize Outreach AB 2024. Tue. February 6, 2024.

   https://www.nobelprize.org/prizes/physics/2021/popular-information/

5. Chen, Xiaosong, and Jingfang Fan (2022): "Opportunities for Complexity Science: the Nobel Prize in Physics 2021", *Physics,* **51** (1), 1-9 (in Chinese).

6. Parisi, Giorgio (2022): "Thoughts on Complex Systems: An Interview with Giorgio Parisi," *Journal of Physics: Complexity*, **3**, 040201.

7. Cugliandolo, Leticia F (2023): "A Scientific Portrait of Giorgio Parisi: Complex Systems and Much More," *Journal of Physics: Complex*, **4**, 011001.

8. Anderson, Phillip W. (1972): "More Is Different," *Science*, **177** (4047), 393-396.

9. Shi, Leilei (2006): "Does Security Transaction Volume Price Behavior Resemble a Probability Wave?" *Physica A,* **366**, 419-436.

10. Holland, John H. (1992): "Complex Adaptive Systems," *Daedalus*, **121** (1), A New Era in Computation

(Winter, 1992), 17-30. https://www.jstor.org/stable/20025416

11. Carmichael, Ted, and Mirsad Hadžikadić (2019): "The Fundamentals of Complex Adaptive Systems," In Ted Carmichael, Andrew J. Collins, and Mirsad Hadžikadić (eds.) *Complex Adaptive Systems, Views from the Physical, Natural, and Social Sciences*, Cham (Switzerland): Springer Nature Switzerland AG, 1-16.

12. Shi, Leilei, Xinshuai Guo, Andrea Fenu, and Bing-Hong Wang (2023): "The Underlying Coherent Behavior in Intraday Dynamic Market Equilibrium," *China Finance Review International*, **13** (4), 568-598.

13. Staddon, John E. R. (2016), *Adaptive Behavior and Learning (2$^{nd}$ Edition)*, Cambridge, London, New York, New Rochelle, Melbourne, Sydney: Cambridge University Press.

14. Pierce, W. David, and Carl D. Cheney (2004): *Behavior Analysis and Learning (3$^{rd}$ Edition)*, Mahwah, New Jersey: Lawrence Erlbaum Associates, Inc., Publishers.

15. Barabási, Albert-László (2016): *Network Science*, Cambridge: Cambridge University Press.

16. Gros, Claudius (2015): *Complex and Adaptive Dynamical Systems (Fourth Edition)*, Springer: Cham Heidelberg New York Dordrecht London.

17. Tao, Terence (2012): "E Pluribus Unum: From Complexity, Universality," *Daedalus (The Journal of the American Academy of Arts & Sciences)*, **141** (3), 23-34.

18. Di, Zengru (2023): "From Phenomenon to Mechanism: The Kepler Era of Social Physics," *Keynote Speaker's Lecture at the Second Conference on Social Physics*, Wuhan (Peking University-Wuhan Institute for Artificial Intelligence), China, November 17-19.

19. Hernández-López, Claudio, Paul Baconnier, Corentin Coulais, Olivier Dauchot, and Gustavo Düring (2024): "Model of Active Solids: Rigid Body Motion and Shape-Changing Mechanisms," *Physical Review Letters*, **132**, 238303.

20. Holland, John H. (2006): "Studying Complex Adaptive Systems," *Journal of Systems Science and Complexity*, **19**, 1-8.

21. Elliott, Thomas J., Mile Gu, Andrew J. P. Garner, and Jayne Thompson (2022): "Quantum Adaptive Agents with Efficient Long-Term Memories," *Physical Review X*, **12**, 011007.

22. Egger, D. J. and F. K. Wilhelm (2014): "Adaptive Hybrid Optimal Quantum Control for Imprecisely Characterized Systems," *Physical Review Letters*, **112**, 240503

23. Einstein, Albert, Boris Podolsky, and Nathan Rosen (1935): "Can Quantum-Mechanical Description of Physical Reality Be Considered Complete?" *Physical Review,* **47** (May 15), 777–780.

24. Schrödinger, Erwin (1935): "Discussion of Probability Relations between Separated Systems," *Mathematical Proceedings of the Cambridge Philosophical Society*, **31**(4), 555-563.

25. He, Qiongyi (2024): "Tests on Complex Quantum Entanglement and Its Applications," *Seminar Reports at Fudan Theoretical Physics*, Video Available at KouShare (April 12), DOI link: https://dx.doi.org/10.12351/ks.2404.0009

26. Colciaghi, Paolo, Yifan Li, Philipp Treutlein, and Tilman Zibold (2023): "Einstein-Podolsky-Rosen Experiment with Two Bose-Einstein Condensates," *Physical Review X*, **13**, 021031.

27. Andrews, M. R, C. G. Townsend, H. J. Miesner, D. S.Durfee, D. M. Kurn, and W. Ketterle (1997): "Observation of Interference between Two Bose Condensates," *Science,* **275**, 637-641.

28. Liu, Wu-Ming, Biao Wu, and Qian Niu (2000): "Nonlinear Effects in Interference of Bose-Einsten Condensates," *Physical Review Letters*, *84* (11), 2294-2297.

29. Zhang, Zhendong, Kai-Xuan Yao, Lei Feng, Jiazhong Hu, and Cheng Chin (2020): "Pattern Formation in a Driven Bose-Einstein Condensate," *Nature Physics*, **16**, 652-656.

30. Schrödinger, Erwin (1928): *Collected Papers on Wave Mechanics (translated from the second German Edition),* London and Glasgow: Blackie, 1–12.

31. Ge, Wei-Kun (2022): "On the Interpretation of the 2022 Nobel Prize in Physics," *Physics*, **51**(12), 821-826 (in Chinese).

32. Popular information (2022): "How Entanglement Has Become a Powerful Tool," *The Nobel Prize in Physics*

*2022*, NobelPrize.org. Nobel Prize Outreach AB 2024. Sat. February 17, 2024.

33. Sun, Chang-Pu (2017): "On Interpretations of Quantum Mechanics," *Physics*, **46** (8), 481-498 (in Chinese).

34. Sun, Chang-Pu (2023): *Open Lectures on Advanced Quantum Mechanics*, Hosted by the Graduate School of China Academy of Engineering Physics, Video Available at KouShare Academic Video Live.

35. Bell, John S. (1964): "On the Einstein, Podolsky, Rosen Paradox," *Physics*, **1**, 195.

36. Aspect, Alain (Translated into Chinese by Xuzong Chen and Zuoyao Chen) (2022): "Bell's Theorem: The Naïve View of an Experimentalist," *Physics*, **51** (12), 827-843.

37. Cui, Lian-Xiang, Kang Xu, Peng Zhang, Chang-Pu Sun (2023): "Quantum Violation of Bell's Inequality and Its Experimental Test—on the Nobel Prize in Physics 2022," *Physics*, **52** (1), 1-17 (in Chinese).

38. Skinner, B. F. (1938): *The Behavior of Organisms: An Experimental Analysis*, New York: Appleton-Century-Crofts.

39. Tversky, Amos, and Daniel Kahneman (1974): "Judgment under Uncertainty: Heuristics and Biases", *Science*, **185**, 1124-1131.

40. Lo, Andrew W. (2017): *Adaptive Markets: Financial Evolution at the Speed of Thought*, Princeton, and Oxford: Princeton University Press.

41. Shiller, Robert J. (1981): "Do Stock Prices Move too Much to be Justified by Subsequent Changes in Dividends?" *The American Economic Review*, **71**, 421-436.

42. Jegadeesh, Narasimhan, and Sheridan Titman (1993): "Returns to Buying Winners and Selling Losers: Implications for Stock Market Efficiency", *Journal of Finance*, **48**, 65-91.

43. Gao, Ya, Xing Han, Youwei Li, Xiong Xiong (2019): "Overnight Momentum, Informational Shocks, and Late Informed Trading in China," *International Review of Financial Analysis*, 66, November, 101394.

44. Yang, Lu (2022): "Last Hour Momentum in the Chinese Stock Market", *China Financial Review International*, **12** (1), 69-100.

45. De Bondt, Werner F. M., and Richard H. Thaler (1985): "Does the Stock Market Overreact?" *Journal of Finance*, Vol. 40 No. 3, pp. 793-805.

46. Münnix, Michael C., Takashi Shimada, Rudi Schäfer, Francois Leyvraz, Thomas H. Seligman, Thomas Guhr, and H. Eugene Stanley (2012): "Identifying States of a Financial Market," *Scientific Reports* 2, 00644.

47. Shi, Leilei, Liyan Han, Yiwen Wang, Yan Piao, Ding Chen, and Chengling Gou (2011): "Market Crowd's Trading Conditioning and Its Measurement", *In Working Paper*, Presentations at the 10th China Economics Annual Conference (2010), 7th Annual Meeting of Chinese Finance Association (2010), 2010 Econophysics Colloquium (Taipei), the 60th Annual Meeting of Midwest Financial Association (2011, USA), 2011 China International Conference in Finance (CICF). Available at SSRN: http://ssrn.com/abstract=1661515

48. Shi, Leilei (2013): "The Volume and Behavior of Crowds", *Automated Trader*, Q2, 90–92, Available at SSRN: http://ssrn.com/abstract=3532838.

49. Elkind, Daniel, Kathryn Kaminski, Andrew W. Lo, Kien Wei Siah, and Chi Heem Wong, (2022): "When Do Investors Freak Out? Machine Learning Predictions of Panic Selling", *Journal of Financial Data Science,* **4** (1), 11–39.

50. Shi, Leilei, Binghong (Bing-Hong) Wang, Xinshuai Guo, and Honggan Li (2021): "A Price Dynamic Equilibrium Model with Trading Volume Weights Based on a Price-Volume Probability Wave Differential Equation", *International Review of Financial Analysis*, **74** (March), 101603.

51. Liang Y., K. N. An, G. Yang, and J. P. Huang (2013): "Contrarian Behavior in a Complex Adaptive System," *Physical Review E*, **87**, 012809.

52. Soros, George (1994): *The Alchemy of Finance,* New York: John Wiley & Sons, Inc., 41-45.

53. Derbes, D. (1996): "Feynman's Derivation of the Schrödinger Equation," *American Journal of Physics*, **64**, 881–884.

54. Zeng, Jinyan (2014): *Quantum Mechanics II* (th fifth eds., pp. 42-47), Beijing: Science China Press.

55. Greenwood, Donald T. (1977): *Classical Dynamics*, Englewood Cliffs, N.J. USA: Prentice-Hall, Inc., 187-213.

56. Everett, Hugh (1957): "Relative State Formulation of Quantum Mechanics," *Review of Modern Physics*, **29**, 454–462.

57. Tang, Pengju, Xiangyu Dong, Wenjun Zhang, Yunhong Li, Xuzong Chen, and Xiaoji Zhou (2020): "Implementation of a double-path multimode interferometer using a spinor Bose-Einstein condensate," *Physical Review A*, **101**, 013612.

58. Bak, Per, Chao Tang, and Kurt Wiesenfeld (1987): "Self-Organized Criticality: An Explanation of 1/$f$ Noise," *Physical Review Letter*, **59** (4), 381-384.

59. Lee, Ryan H., Erwin A. B. Mulder, and Jonathan B. Hopkins (2022): "Mechanical Neural Networks: Architected Materials that Learn Behaviors," *Science Robotics*, **7** (71), 19.

## Data availability statement

This article is purely theoretical. It has evidencing data support from previous empirical studies in finance [9] [47] [50] and experimental supports in quantum mechanics [25-27]. The datasets in finance are available in supplemental files.

## Acknowledgments

The authors appreciate discussions with Wu-Ming Liu, Xuzong Chen, Qiongyi He, Zhengwei Zhou, Yinghua Zhang, Zhigang Zheng, Linyuan Lv, Tianlong Fan, Haotian Shi, Wencheng Zhang, Huaiyu Wang, Haiqing Lin, Yihuo Ye, Zejun Ding, Stanisław Drożdż, Everett X. Wang, Huijie Yang, Song Chen, Tianpei Wang, Yihuo Ye, Zhong Wang, and the conference participants from the 10$^{th}$ International Symposium on Cold Atom Physics (Shanghai, June 27-29, 2024 forthcoming), 21th Quantum Optics Conference (Fuzhou, August 9-12, 2024 forthcoming), the 8$^{th}$ Conference on Systems Science of China (Wuhu, May 17-19, 2024), the 7$^{th}$ International Conference on Physics, Mathematics and Statistics (Yichang, May 23-25, 2024), the 2$^{nd}$ China Social Physics Forun (Peking University-Wuhan Institute for Artificial Intelligence), ArtInHCI 2023 International Conference on Artificial Intelligence and Human-Computer (online conference), International Conference on Econophysics (Shanghai, 2023), and the 7$^{th}$ Conference on Systems Science of China (Chongqing, 2023). Bing-Hong Wang, Jiuchang Wei, and Guocheng Wang thank the funding from the National Natural Science Foundation of China (Grant No: 71874172, 72293573 and 72003007), respectively. The authors are responsible for all omissions and errors.

## Funding

Bing-Hong Wang, Jiuchang Wei, and Guocheng Wang thank the funding from the National Natural Science Foundation of China (Grant No: 71874172, 72293573 and 72003007), respectively.

## Author contributions statement

Shi contributed to half of this work, and the others contributed equally. Specifically

Conceptualization: LS, XG, JW, WZ, GW, and B.-H.W.

Methodology: LS, JW and B.-H.W.

Investigation: LS and XG.

Supervision: LS, B.-H.W., XG, JW, and GW.

Writing—original draft: LS.

Writing—review & editing: LS, XG, WZ.

All authors reviewed the manuscript.

## Conflict of interest

The authors declare that they have no conflict of interest.